# Electron interaction with the angular momentum of the electromagnetic field

R. F. O'Connell[*]

Department of Physics and Astronomy, Louisiana State University, Baton Rouge, LA 70803-4001, USA

We give a simple derivation and expansion of a recently proposed new relativistic interaction between the electron and the angular momentum of the electromagnetic field in quantum electrodynamics (QED). Our derivation is based on the work of Moller, who pointed out that, in special relativity, a particle with spin must always have a finite extension. We generalize Moller's classical result to the quantum regime and show that it leads to a new contribution to the energy, which is the special relativistic interaction term. In addition, we show that all spin terms arising from the Dirac equation may be obtained by this method.

A new relativistic interaction term in QED which, in particular, couples the angular momentum of light and the election spin was initially proposed on the basis of symmetry arguments [1] and verified in a complex calculation involving the Dirac equation and the Foldy-Wouthuysen transformation [2]. In many ways, this is a surprising result, since it did not appear explicitly in the avalanche of work on QED following Dirac's work. Moreover, since the orbital momentum properties of light are now of much interest, we are motivated to provide a simple derivation and physical explanation. In addition, we show how this result may be written in a revealing manner as an interaction between the magnetic moment of the electron and the magnetic moment of the electromagnetic field. In addition, we show that all spin terms arising from the Dirac equation may be obtained by this method.

Our derivation is based on the work of Moller [3], who pointed out that, in special relativity, a particle with structure and spin S (its angular momentum vector in the rest system $K^{(0)}$) must always have a finite extension and that there is a "... difference $\Delta r$ between simultaneous positions of the center of mass in its own rest system $K^{(0)}$ and system K (obtained from $K^{(0)}$ by a Lorentz transformation with velocity $v$)...". We generalize Moller's result to the quantum regime and derive what we call "hidden energy", which is the new relativistic interaction term, in addition to the usual spin-orbit interaction term.

We refer to $\Delta \boldsymbol{r}$ as hidden position and it is given by [3]

$$\Delta \boldsymbol{r} = \frac{S \times v}{mc^2} \quad . \tag{1}$$

As we have previously shown [4], if we take the time derivative of $\Delta \boldsymbol{r}$, we obtain the hidden velocity $\Delta \boldsymbol{v}$. Next, we work to order $c^{-2}$ (as in Refs. 1 and 2), so that we can neglect the very small second order terms, which arise in the relation between the hidden velocity and the hidden momentum as well as in the time dependence of the spin, to obtain an expression for the hidden momentum

$$\Delta \mathbf{p} = \frac{S \times a}{c^2} = \frac{S \times F}{mc^2} \quad , \tag{2}$$

where

$$\mathbf{p} = m\mathbf{v}, \quad (3)$$

$\boldsymbol{a}$ is the acceleration, and **F** is the external force acting on the particle. The existence of hidden momentum is now a well-documented phenomenon in electromagnetism [5] but it also plays an important role in the general relativistic treatment of spinning particles [6,7]. However, it should be emphasized that equation (2) is not unique in the sense that it depends on the coordinate system chosen and, in the coordinate system used in the derivation of the Dirac equation [6,7], $\Delta \boldsymbol{p}$ is – ½ of the result given in (2). In essence this is related to the fact that, in special relativity, spin is described by a second-rank tensor $S_{\alpha\beta} = -S_{\beta\alpha}$ or by an axial 4-vector $S^{\mu}$ which reduces to the 3-vector **S** when the particle velocity is zero.

Explicitly,

$$S_\alpha \equiv \frac{1}{2} \epsilon_{\alpha\beta\sigma\tau} S^{\beta\sigma} U^\tau, \quad (4)$$

and

$$S^{\alpha\beta} = \epsilon^{\alpha\beta\sigma\tau} S_\sigma U_\tau. \quad (5)$$

The relation between $S_{\alpha\beta}$ and $S^{\mu}$ is not unique [8] but depends on the choice of the so-called spin supplementary condition, which in turn depends on the coordinate system chosen. Popular choices are $S^{\alpha\beta}U_\beta = 0$ and $S^{\alpha\beta}p_\beta = 0$ where $U_\beta$ and $p_\beta$ are 4-velocity and 4-momentum vectors. However, as is clear from Ref. 6, these choices are not suitable for treating an accelerating particle, and, instead, we introduced a new supplementary condition, given in (6) below, which we showed to be consistent with Dirac's coordinate system. In addition, we showed that our new choice was the basis for explaining the difference in results obtained by Schiff [9] and Ref. 6. In fact, Schiff's choice corresponded to the $\Delta \boldsymbol{r}$ given in (2) whereas Barker and O'Connell [7], using Dirac's coordinate system obtained $-\Delta \boldsymbol{r}/2$. The factor of 2 provides an important clue when we recall the famous factor of 2 that arose from Thomas precession (rotation) after he took into account the extra Lorentz transformation arising from acceleration. Since acceleration is always playing a key role, the popular choices mentioned above refer to non-inertial frames of reference and, as we showed in [7], Dirac's choice (essentially what Jackson[10] refers to as the nonrotating frame) actually corresponds to choosing an inertial frame given by

$$2S^{i4} + S^{ij} U_j = 0, \quad (6)$$

which incorporates the Thomas spin-orbit contribution [10]. On the other hand, Moller's choice $S^{\alpha\beta}U_\beta = 0$ corresponds to choosing the rest frame of the electron (a non-inertial frame) and thus the quantum generalization does not include the Thomas contribution.

Furthermore, the results of Ref. 6 were based on a potential derived from a one-graviton exchange between two Dirac particles, from which the classical spin precession results were obtained by letting ½ $\hbar \boldsymbol{\sigma} \rightarrow \boldsymbol{S}$. Thus, it is clear that, in the Dirac coordinate system, the quantum generalizations of equations (1) and (2) are

$$\Delta \boldsymbol{r} = -\frac{\hbar}{2} \frac{\boldsymbol{\sigma} \times \boldsymbol{v}}{2mc^2}, \quad (7)$$

and

$$\Delta \boldsymbol{P} = -\frac{\hbar}{2} \frac{\boldsymbol{\sigma} \times \boldsymbol{a}}{2c^2}, \quad (8)$$

where **P** is the canonical momentum, given by

$$P = p + eA, \quad (9)$$

and where **A** is the vector potential of the electromagnetic field.

Here, we wish to apply the above expressions to incorporate spin effects into the non-relativistic quantum Hamiltonian for a point structureless particle, in a potential $V(r)$, with charge e, interacting with the electromagnetic field, given by [11]

$$H = \frac{1}{2m}(P - eA)^2 + V(r). \quad (10)$$

We also use the Coulomb (Radiation) gauge, a common choice [11]. Thus, introducing spin effects via the above equations implies that, to order $c^{-2}$,

$$\Delta H = -eA \cdot \frac{\Delta P}{m} + P \cdot \frac{\Delta P}{m} + \Delta V(r). \quad (11)$$

Thus,

$$\Delta H = -eA \cdot \frac{\Delta P}{m} + P \cdot \frac{\Delta P}{m} - \frac{dV}{dr}(\Delta r \cdot \hat{r}). \quad (12)$$

Using (7), we see that

$$-\frac{dV}{dr}(\Delta r \cdot \hat{r}) = \frac{\hbar}{4mc^2} \frac{1}{r} \frac{dV}{dr} \sigma \cdot (r \times v)$$

$$= \frac{1}{2m^2c^2} \frac{1}{r} \frac{dV}{dr}(S \cdot L), \quad (13)$$

where **L** is the angular momentum of the electron. There is a very familiar result [10,11] for the spin-orbit interaction.

Next, turning to the first term in (12) and using (8), we obtain

$$\Delta H_1 \equiv -eA \cdot \Delta P/m = \frac{e\hbar}{4mc^2} A \cdot (\sigma \times a). \quad (14)$$

Taking $a = (eE/m)$, where **E** is the electric field (but noting that the external force may be generalized to include magnetic contributions) and rearranging the cross product, we obtain our basic result

$$\Delta H_1 = \frac{e^2\hbar}{4m^2c^2} \sigma \cdot (E \times A). \quad (15)$$

The expectation value of $\Delta H_1$ is the hidden energy. It will be recognized that this term is exactly the same as the $H_{AME}$ interaction term of Mondal et al., which all their many applications are based on. Further physical insight is obtained by recognizing that $(E \times A)$ is the angular momentum density of the electromagnetic field [12]. Thus, the corresponding magnetic moment $M_{em}$ is $(e/2mc)(E \times A)$ and recalling that the magnetic moment of the electron $M_e$ is $(e\hbar/2mc)\sigma$, we see that

$$\Delta H_1 = M_{em} \cdot M_e. \quad (16)$$

In summary, Moller's special relativistic hidden position leads to hidden momentum and hidden energy, whose quantum version leads to the usual spin-orbit term as well as a new expression for the interaction of the electron spin with the angular momentum of the electromagnetic field. Finally, using (8), we see that the second term in (11) may be written as

$$\Delta H_2 = -\hbar \sigma \cdot (a \times P)/4mc^2, \quad (17)$$

a result obtained by Hehl and Ni [13] in a tour-de-force calculation necessitating the use of three successive Foldy-Wouthuysen transformations. In the particular case where $a = eE/m$, we obtain the well-known result [14]

$$\Delta H_2 = -e\hbar \sigma \cdot (E \times P)/4m^2c^2. \quad (18)$$

In conclusion, we have shown that all spin terms arising from the Dirac equation may be obtained from a quantum generalization of Moller's observation and, in particular, we have given an explicit expression for the electron interaction with the angular momentum of the electromagnetic field.

This work was partially supported by the National Science Foundation under grant no. ECCS-1125675.

The author appreciates a discussion on [12] with Professor A. Zangwill and comments from Professor P.M. Oppeneer. In addition, Dr. Mondal and Professor Oppeneer pointed out that my initial draft of the paper omitted the **P. ΔP** term. The author also appreciates a communication from Professor Hehl who informed me of his work with Ni in [13], which had already identified this new inertial spin-orbit coupling term. He is particularly grateful to Dr. Paul Werbos, former NSF program manager, who was always receptive to the pursuit of projects "outside the box" which had the potential to produce innovative results.

________________

- oconnell@phys.lsu.edu

[1] A. Raeliarijaona et al, Phys. Rev Lett. **110**, 137205 (2013).
[2] R. Mondal et al, Phys. Rev. B **92**, 100402 (2015).
[3] C. Moller, Commun. Dublin Inst. Adv. Stud., **A5**, 1 (1949); ibid. "The Theory of Relativity," 2nd ed. (Oxford University Press, 1972), p. 172.
[4] R.F.O' Connell in "Equations of Motion in Relativistic Gravity", edited by D. Puetzfeld et al, Springer "Fundamental Theories of Physics" **179** (Springer 2015).
[5] D.J. Griffiths, Am. J. Phys. **80**, 7 (2012).
[6] B.M. Barker and R.F. O'Connell, Phys. Rev. D**2**, 1428(1970).
[7] B.M. Barker and R.F. O' Connell, Gen. Relat. Gravit.**5**, 539 (1974).
[8] R.F.O' Connell, "Rotation and Spin in Physics," in General Relativity and John Archibald Wheeler, ed. I. Ciufolini and R. Matzner, (Springer, 2010).
[9] L.I. Schiff, Proc. Natl. Acad. Sci. **46**, 871 (1960).
[10] J.D. Jackson. *Classical Electrodynamics*, Third Edition (Wiley, New York, 1998), Sections 11.8 and 11.11.
[11] R. Shankar, *Principles of Quantum Mechanics*, 2nd edition. (Plenum Publishers, New York, 1994), pps.91 and 500.
[12] A. Zangwill, *Modern Electrodynamics* (Cambridge U.Press, 2013), Problem 16.7, p.578
[13] F.W.Hehl and W.T. Ni, Phys.Rev.D 42, 2045 (1990).
[14] J.D.Bjorken and S.D.Drell, "Relativistic Quantum Mechanics", (McGraw-Hill Press,1964),p.51,Eq. (4.5)